\documentclass[12pt]{amsart}
\usepackage{amsmath, amssymb}

\theoremstyle{plain}
\newtheorem{theorem}{Theorem}[section]

\newtheorem{problem}[theorem]{Problem}

\theoremstyle{definition}

\theoremstyle{remark}

\numberwithin{equation}{section}

\DeclareMathOperator{\edges}{edges}

\def \R {\mathbb{R}}

\def \E {\mathbb{E}}
\def \P {\mathbb{P}}

\def \e {\varepsilon}

\def \k {\kappa}
\def \l {\lambda}

\def \s {\sigma}

\def \< {\langle}
\def \> {\rangle}
\def \^ {\widehat}

\def \Span {{\rm span}}

\def \conv {{\rm conv}}

\def \lmax {\l_{{\rm max}}}
\def \lmin {\l_{{\rm min}}}

\begin{document}
\title[]{Some problems in asymptotic convex geometry
  and random matrices motivated by numerical algorithms}

\author{Roman Vershynin}

\thanks{Supported by the Alfred P.~Sloan Foundation
  and by NSF DMS grant 0401032.}

\address{Department of Mathematics,
   University of California,
   Davis, CA 95616, USA}
\email{vershynin@math.ucdavis.edu}

\date{\today}

\begin{abstract}
  The simplex method in Linear Programming motivates
  several problems of asymptotic convex geometry. 
  We discuss some conjectures and known results in two related directions --
  computing the size of projections of high dimensional polytopes
  and estimating the norms of random matrices and their inverses.
\end{abstract}

\maketitle

\section{Asyptotic convex geometry and Linear Programming}
%---------------------------------------------------------

Linear Programming studies the problem of maximizing a 
linear functional subject to linear constraints. 
Given an {\em objective vector} $z \in \R^d$
and {\em constraint vectors} $a_1,\ldots,a_n \in \R^d$, 
we consider the linear program

\begin{equation*}                       \tag{LP}
\begin{aligned}
  &\text{maximize } \< z, x \>  \\
  &\text{subject to } \< a_i, x\> \le 1, \quad i=1,\ldots,n.
\end{aligned}
\end{equation*}

This linear program has $d$ unknowns, represented by $x$,
and $n$ constraints. Every linear program can be reduced to this form
by a simple interpolation argument \cite{V}.
The {\em feasible set} of the linear program is the polytope 
$$
P := \{ x \in \R^d :\; \< a_i, x\> \le 1, \quad i=1,\ldots,n \}.
$$
The solution of (LP) is then a vertex of $P$.
We can thus look at (LP) from a geometric viewpoint: 
\begin{quote}
  for a polytope $P$ in $\R^d$ given by $n$ faces,
  and for a vector $z$,
  find the vertex that maximizes the linear functional $\< z, x\> $.
\end{quote}

The oldest and still the most popular method to solve this problem is 
the {\em simplex method}. 
It starts at some vertex of $P$ and generates a walk on the edges
of $P$ toward the solution vertex. 
At each step, a {\em pivot rule} determines a choice of the next vertex;
so there are many variants of the simplex method with different pivot rules.
(We are not concerned here with how to find the initial vertex, 
which is a nontrivial problem in itself).

\subsection{The shadow-vertex pivot rule and sections of polytopes}
%.....................................................................

The most widely known pivot rule maximizes the objective function $\< z, x\> $
over the neighboring vertices. The resulting walk on the vertices is
defined iteratively, and thus is usually hard to analyze. 
An alternative {\em shadow-vertex} pivot rule \cite{GS} 
defines a walk on the polytope $P$ as a preimage of a projeciton of $P$.
The resulting walk can be desciribed in a non-iterative way, so one 
hopes to analyze it with the methods of asymptotic convex geometry.

Suppose we know a solution $x_0$ of (LP) for some other objective vector
$z_0$. The shadow-vertex simplex method interpolates between $z_0$ and $z$ 
by computing the solutions of (LP) for all $z'$ in the plane
$$
E = \Span( z_0,z ).
$$

From a geometric viewpoint, we consider the orthogonal projection 
$Q(P)$ of the feasible polytope $P$ onto $E$.
It is easily checked that the vertices $x_0$ and $x$ of $P$ will be preserved
by the projection: $Q(x)$ and $Q(x_0)$ will be vertices of the polygon $Q(P)$.

The shadow-vertex simplex method thus computes the vertices of the polygon $Q(P)$
one by one, starting from $Q(x_0)$ and ending with $Q(x)$. So at the end it 
outputs $x$, which is the solution of (LP). One can express the computation of 
$Q(x)$ as a pivot rule, and check that each next vertex can be computed
in polynomial time. The resulting walk on the polytope $P$ is therefore 
the preimage of the vertices of the polygon $Q(P)$ under the projection $Q$. 

It will be convenient to work in the dual setting. The polar of $P$ is 
$$
K := P^\circ = \{ x \in \R^d :\; \< x,y \> \le 1 \text{ for all } y \in P \}
= \conv(0,a_1,\ldots,a_n)
$$
and the polar of the projection $Q(P)$ is the section $K \cap E$.
The length of the walk in the shadow-vertex simplex method is thus 
bounded by the size (the number of edges) of the polygon $K \cap E$.

\subsection{Complexity of the simplex method and the size of sections}
%....................................................................

The running time of the simplex method is proportional to the length of the walk
on the edges of $P$ it generates. 
Hirsch's conjecture states that every polytope 
$P$ in $\R^d$ with $n$ faces has {\em diameter} at most $n-d$. 
The diameter is the maximum of the shortest walk on the edges between any pair of the vertices. The best known bound on the diameter is 
$n^{\log_2 d + 2}$ due to Kalai and Kleitman \cite{KK}.

For every known variant of the simplex method, an example of (LP) is known
for which the length of the walk on $P$ is not polynomial in $n$ and $d$.\footnote{Recently, a prandomized polynomial time 
pivoting algorithm for (LP) was found by Kelner and Spielman \cite{KS}. 
However, their algorithm generates a walk on some other polytope related to (LP)
and not on $P$.}
For the the classical (maximizing) pivot rule, such an example was first
constructed by Klee and Minty \cite{KM}: on a certain deformed cube, 
the walk visits each of the $2^d$ vertices \cite{KM}.

Similar pessimistic examples are known for the the shadow-vertex simplex method: 
the size of the planar section $K \cap E$
that bounds the length of the walk is in general exponential in $n,d$. 
This follows for example from the seminal construction in semidefinite
programming by Ben-Tal and Nemirovski \cite{BN}, 
which yields a polytope $K$ and a plane $E$ such that the section $K \cap E$ 
is an approximation of the circle with error {\em exponentially small} in $n,d$.

\begin{problem}[Sections of polytopes]			\label{p: sections}
  Let $K$ be a polytope in $\R^d$ with $n$ vertices,
  and $E$ be a two-dimensional subspace of $\R^d$.
  Estimate the size (the number of edges) of the polygon $K \cap E$. 
  Under what conditions on $K$ and $E$ is this number polynomial in $n,d$?
\end{problem}

This problem is somewhat opposite to the typical problems of the 
asymptotic convex geometry, whose ideal would be to produce the most round
section (fine approximation to a cicrle). In Problem~\ref{p: sections}
our ideal is a section with fewest edges, thus farthest from the circle. 
From the viewpoint of the simplex method, ``round'' polytopes have high complexity, while polytopes with fewest faces have low complexity.

\subsection{Smoothed analysis and randomly perturbed polytopes}  \label{s: sa}
%................................................................

Despite the known examples of exponentially long walks, on most problems that 
occur in practice the simplex algorithm runs in polynomial and even linear time. 
To explain this empirical evidence, the {\em average analysis} of the 
simplex method was developed in the eighties, where the (LP) was drawn 
at random from some natural distribution and the expected size of the walk
was shown to be polynomial in $n,d$ 
\cite{Bor, Sma1, Sma2, Meg86, Hai83, Adl83, Tod86, AM85, AKS87}.

In particular, Haimovich showed (\cite{Hai83}, see \cite{Sch}, Section 11.5)
that if one chooses the directions of the inequalities in (LP) 
uniformly at random as $\le$ or $\ge$, 
then the expected length of the walk in the shadow-vertex simplex method
is at most $d/2$. Note that the size does not depend on the number of inequalities
$n$.

However, reversing inequalities is hard to justify in practice.
Spielman and Teng \cite{ST Congress} proposed to replace average analysis
by a finer model, which they called {\em smoothed analysis},
and where the random inputs are replaced by slight {\em random perturbations}
of arbitrary inputs. Smoothed analysis thus interpolates between
the worst case analysis (arbitrary inputs) and the average analysis
of Smale (random inputs). 

Spielman and Teng \cite{ST} first showed that the shadow-vertex simplex method
has polynomial smoothed complexity. If the polytope $K$ is randomly perturbed,
then its section $K \cap E$ will have an expected polynomial size
(which in turn bounds the length of the walk in the simplex method).
Their result was improved in \cite{DS} and in \cite{V}, 
and the current best bound is as follows:

\begin{theorem}\cite{V}				\label{t: section bound}
  Let $a_1,\ldots,a_n$ be independent Gaussian vectors in $\R^d$
  with centers of norm at most $1$, and whose components have 
  standard deviation $\s \le 1/6\sqrt{d \log n}$.
  Let $E$ be a plane in $\R^d$. Then the random polytope
  $K = \conv(a_1,\ldots,a_n)$ satisfies
  \begin{equation}                                  \label{section bound}
    \E \, |\edges (K \cap E)| \le C d^3 \s^{-4},
  \end{equation}
  where $C$ is an absolute constant.
\end{theorem}

The prior weaker bound of Spielman and Teng \cite{ST} was $C n d^3 \s^{-6}$;
the subsequent work of Deshpande and Spielman \cite{DS} 
improved upon the exponent of $d$ but doubled the exponent of $n$.

\medskip

Theorem~\ref{section bound} shows that the expected size of the 
section $K \cap E$ is {\em polylogarithmic} in $n$,
while the previous bound were polynomial in $n$. 
Going back to the pre-dual polytope $P$, this indicates that random 
perturbations of the polytopes create short walks between any two {\em given} 
vertices.Note that for large $n$, this polylogarithmic bound becomes 
better than the bound $n-d$ in Hirsch's conjecture.
\medskip

Theorem~\ref{section bound} provides a solution to Problem~\ref{p: sections}
for a randomly perturbed polytope and a fixed subspace.
A seemingly harder problem, which is still open, is for an {\em arbitrary 
polytope and a randomly perturbed subspace}. This version 
would be significant for the analysis of the simplex method, 
because it allows one to leave 
the constraints intact and only perturb the objective function.

\medskip

Another open problem is to estimate the diameter of 
randomly perturbed polytopes, rather than all polytopes as in 
Hirsch's conjecture.

\begin{problem}[Spielman-Teng \cite{ST}]
  Let $K$ be a perturbed polytope as in Theorem~\ref{t: section bound}. 
  Estimate the expected diameter of $P = K^\circ$.
  Is it always polynomial in $n$, $d$ and $\s$?
  Perhaps even polylogarithmic in $n$?
\end{problem}

Finally, no analog of Theorem~\ref{t: section bound}
is known for {\em bounded perturbations}, i.e. for 
for $a_i = \bar{a}_i + \s \theta_i$, where $\bar{a}_i$ are arbitrary 
fixed vectors of norms at most $1$ and $\theta$ are independent vectors
chosen from $\{-1,1\}^d$ or from $[-1,1]$ uniformly at random.
Such {\em bounded smoothed analysis} is a common 
model for roundoff errors when real numbers are represented as 
binary numbers in computers \cite{GN}.

\subsection{Nondegeneracy of faces and invertibility of random matrices}
						\label{s: nondegeneracy}
%...........................................................................

The approach to Theorem~\ref{t: section bound} developed by Spielman and Teng
\cite{ST} is based on the intuition that most faces of $K$ should be 
non-degenerate simplices (e.g. they have inscribed balls of polynomial radii). 
If the plane $E$ intersects such a nondegenerate simplex $F$, the length
of the interval $E \cap F$ is likely to be polynomially big (if the plane
intersects a simplex, it is likely to pass through its ``bulk'' rahther than
touch the boundary only). 

On the other hand, with high probability all vectors in the the perturbed 
polytope $K$ have norms $O(\log n)$. (Its vertices are Gaussian perturbations
of $n$ vectors of norm at most $1$). Therefore, the perimeter of the polygon
$K \cap E$ can be at most $O(\log n)$. Since all edges $E \cap F$ of this polygon
are polynomially big, we conclude that are at most polynomially many edges, 
as desired.

\medskip

There are several places where this approach breaks down or is not known to 
succeed. One such problem is the non-degeneracy of the faces. The
nondegeneracy of a simplex $S$ is usually quantified with the smallest singular 
value of the matrix $A$ that realizes the change of the basis from the standard 
simplex to $S$. For the polytope $K = \conv(a_1,\ldots,a_n)$, 
each face is a simplex with vertices $(a_i)_{i \in I}$ for some 
$d$-element subset $I \subset \{1,\ldots,n\}$. If the vertices $a_i$ are 
Gaussian as in Theorem~\ref{t: section bound}, the change of basis 
is a $d \times d$ matrix with Gaussian independent entries. 
Thus we need the random Gaussian matrices to be far from being singular.
Quantitative theory of invertibility of random matrices is the subject 
of the next section.

\section{Invertibility of random matrices}
%-----------------------------------------

For a one-to-one linear operator $A : X \to Y$ between two normed spaces
$X$ and $Y$, two quantities are central in functional analysis: 
the norm $\|A\|$ and the norm of the inverse $\|A^{-1}\|$. 
If the operator is not onto, then the inverse norm is computed for the 
restriction of $A$ onto its image; so we identify $\|A^{-1}\|$ with
$\|(A|_{A(X)})^{-1}\|$.
Thus 
$$
\|A\| = \sup_{x \in X:\; \|x\| = 1} \|Ax\|, \qquad
\frac{1}{\|A^{-1}\|} = \inf_{x \in X:\; \|x\| = 1} \|Ax\|.
$$
The operator $A$ can be viewed as realizing an {\em embedding} 
of the space $X$ into the space $Y$, and the product $\|A\| \|A^{-1}\|$
is the {\em distortion} of the embedding (see \cite{IM}).

The canonical example is when both $X$ and $Y$ are finite dimensional 
Euclidean spaces, say $X = \R^k$, $Y = \R^n$, where we identify the 
linear operator $A$ with its $k \times n$ matrix. 
The {\em singular values} of $A$ are the eigenvalues of $|A| = \sqrt{A^*A}$,
the largest and the smallest singular values being 
$$
\lmax(A) = \|A\|, \qquad \lmin(A) = \frac{1}{\|A^{-1}\|}.
$$
In the numerical linear algebra and scientific computing literature, 
the distortion 
$$
\qquad \k(A) =  \lmax(A) / \lmin(A) = \|A\|\|A^{-1}\|
$$
is commonly called the {\em condition number} of $A$.

\medskip

We are interested estimating these quantities for random matrices $A$. 
For one reason, random matrices sometimes provide an intuition for what to 
expect in practice; we saw such reasoning about average analysis and smoothed
analysis of the simplex method in the previous section. 
Random linear operators with controllable distortion (or their adjoints)
also serve as handy tools in most randomized
constructions in geometric funcitonal analysis \cite{DS}, 
geometric algorithms in theoretical computer science \cite{Vem1, Vem2},
compressed sensing in information processing (\cite{D}, \cite{CS}),
vector quantization \cite{LV} and some other fields.

\subsection{Gaussian matrices}
%..............................

We start from the simplest case of a {\em Gaussian matrix}, those whose
entries are i.i.d. standard normal random variables. 
The asymptotics of the largest and the smallest singular values is 
well understood in this case: for a $n \times d$ Gaussian matrix $A$
with $n \ge d$, one has
$$
\lmax(A) \approx \sqrt{n} + \sqrt{d}, \quad
\lmin(A) \approx \sqrt{n} - \sqrt{d}
 \quad \text{with high probability.}
$$
There is a long history of such asymptotic results. 
In particular, the largest and the smallest singular values converge 
almost surely to their corresponding values above 
as the dimension $n$ grows to infinity
and the aspect ratio $d/n$ converges to a constant, see \cite{DS}. 
A sharp nonasymptotic result -- for every fixed $n$ and $d$ --
follows from Gordon's inequality (see \cite{DS}):
$$
\sqrt{n} - \sqrt{d}
\le \E \lmin(A) \le \E \lmax(A)
\le \sqrt{n} + \sqrt{d}.
$$
Combining with the concentration of measure inequality, one
deduces a deviation bound \cite{DS}: for every $t > 0$, 
with probability at least $1 - 2e^{-t^2/2}$ one has
\begin{equation}				\label{conc}
  \sqrt{n} - \sqrt{d} - t
  \le \lmin(A) \le \lmax(A)
  \le \sqrt{n} + \sqrt{d} + t.
\end{equation}

\medskip

Note that the lower bounds become meaningless for {\em square Gaussian matrices},
those with $n=d$. Yet this case is central in some applications:
as we saw in Section~\ref{s: nondegeneracy}, the square matrices determine 
the faces of the polytope $K$ in linear programming, and nondegeneracy of 
such a face translates into a lower bound for $\lmin(A)$.
To guess the order of the smallest singular value,
note that for an $(n-1) \times n$ matrix, the lower bound is 
$\sqrt{n} - \sqrt{n-1} \sim n^{-1/2}$. 

Von Neumann and his associates, who used random matrices as test matrices
for their algorithms, indeed speculated that for random square matrices
one should have
\begin{equation}                \label{whp}
  \lmin(A) \sim n^{-1/2} \quad \text{with high probability}
\end{equation}
(see \cite{vN}, pp.~14, 477, 555.)
In a more precise form, this estimate was conjectured by Smale \cite{S 85} 
and proved by Edelman \cite{E 88} for Gaussian matrices: 
for every $\e \ge 0$, one has
\begin{equation}                \label{tail}
  \P \big( \lmin(A) \le \e n^{-1/2} \big) \sim \e.
\end{equation}
In particular, the smallest singular value is not concentrated: 
the mean and the standard deviation of $n^{1/2} \lmin(A)$ are both 
of the order of a constant. This is very different from the behavior
of the largest singular value, which by \eqref{conc}
is tightly concentrated around its mean.

An elegant argument by Sankar, Spielman and Teng \cite{SST}
generalizes \eqref{tail} for random Gaussian perturbations of 
an arbitrary matrix, i.e. for the smoothed analysis setting
of Theorem~\ref{t: section bound}:

\begin{theorem}[Sankar, Spielman and Teng \cite{SST}]	\label{SST}
  Let $A$ is an $n \times n$ matrix with independent Gaussian random 
  entries (not necessarily centered), each of variance $\s^2$. 
  Then, for every $\e \ge 0$, one has
  $$
  \P \big( \lmin(A) \le \e n^{-1/2} \big) \le C\e/\s,
  $$
  where $C = 1.823$.
\end{theorem}

In applications for random polytopes such as in Section~\ref{s: nondegeneracy},
we need {\em all faces} to be nondegenerate, thus all $d \times d$ submatrices
of a random $n \times d$ Gaussian matrix (whose rows are the constraint vectors
$a_1,\ldots,a_n$ from (LP)) be nicely invertible. 
This motivates the following problem: 

\begin{problem}
  Let $A$ be an $n \times d$ Gaussian matrix 
  (with i.i.d. standard normal entries, 
  or, more generally, as in Theorem~\ref{SST}).
  Estimate the expected minimum of the smallest singular values of all
  all $d \times d$ submatrices of $A$.
\end{problem} 

In particular, if $n = O(d)$, we want this minimum to be polynomially small
rahter than exponentially small in $d$.

\subsection{General matrices with i.i.d. entries}
%................................................

Most problems we discussed become much harder once Gaussian matrices
are replaced with other natural matrices with i.i.d. entries. 
Nevertheless, understanding of discrete matrices, whose entries
can take finite set of values, is important in applications
such as in numerical algorithms, which can only deal with discrete values.
A survey on random discrete matrices was recently written by Vu \cite{Vu}.

Asymptotic theory of random matrices has developed to a point where 
the behavior of the largest singular value is well understood. 
Suppose $A$ is an $n \times d$ matrix with i.i.d. centered entries, 
which have variance $1$. Then the finiteness of the {\em fourth moment}
of the entries is necessary and sufficient that  
\begin{equation}				\label{lmax asymptotic}
  \lmax(A) \to \sqrt{n} + \sqrt{d} \qquad \text{almost surely} 
\end{equation}
as the dimension $n$ grows to infinity
and the aspect ratio $d/n$ converges to a constant \cite{YBK}.
A similar statement holds for the convergence in probability, 
and with a slightly weaker condition than the fourth moment \cite{Si}.

Under a much stronger {\em subgaussian} moment assumption, 
which still holds discrete and gaussian random variables, 
a parallel non-asymptotic result is known (for all finite $n$ and $d$).
A random variable $\xi$ is called {\em subgaussian} if its tail is
dominated by that of the standard normal random variable:
there exists a constant $B > 0$ such that
\begin{equation}   \label{subgaussian}
  \P (|\xi| > t) \le 2 \exp(-t^2/B^2) \qquad \text{for all $t > 0$}.
\end{equation}
The minimal $B$ here is called the {\em subgaussian moment}.\footnote{
  In the literature in geometric functional analysis, the subgaussian
  moment is often called the $\psi_2$-norm.} 
Gaussian random variables and all bounded random variables, in
particular the symmetric $\pm 1$ random variable, are examples of
subgaussian random variables. Inequality \eqref{subgaussian} is
often equivalently stated as a moment condition
\begin{equation}    \label{subgaussian moments}
   (\E |\xi|^p )^{1/p} \le C B \sqrt{p} \qquad \text{for all $p \ge 1$},
\end{equation}
where $C$ is an absolute constant.

The following non-asymptotic result follows
from a more general result proved by Klartag and Mendelson
(\cite{KMe}, Theorem~1.4) with constant probability, 
which was later improved by Mendelson, Pajor and Tomczak-Jaegermann 
(\cite{MPT}, Theorem~D) to an exponential probability:

\begin{theorem}[\cite{KMe, MPT}]
  Let $A$ be an $n \times d$ matrix $(n \ge d)$ with i.i.d. centered 
  subgaussian entries with variance $1$.
  Then, with probability at least $1 - C e^{-m}$, one has
  $$
  \sqrt{n} - C \sqrt{d} 
  \le \lmin(A) \le \lmax(A)
  \le \sqrt{n} + C\sqrt{d},
  $$
  where $C$ depends only on the subgaussian moment of the entries.
\end{theorem}

This estimate approaches the sharp asymptotic bound \eqref{lmax asymptotic} 
for very tall matrices (for small aspect ratios $d/n$). 
However, the lower bound becomes useless for the aspect ratios above 
some constant, and in particular says nothing about square matrices.

\begin{problem}					\label{p: lmin}
  Let $A$ be an $n \times d$ matrix $(n \ge d)$ with i.i.d. centered 
  subgaussian entries with variance $1$.
  Is it true that with high probability one has 
  $$
  \lmin(A) \ge c (\sqrt{n} - \sqrt{d}),
  $$
  where $c>0$ depends only on the subgaussian moment of the entries?
\end{problem}

In a positive direction, lower bounds valid for all aspect ratios 
$y:= d/n < 1$ were proved by Litvak, Rudelson, Pajor and Tomczak-Jaegermann
\cite{LPRT} with an exponential dependence on $1-y$, and improved to a linear
dependence by Rudelson \cite{R}.

\medskip

Nevertheless, even the positive solution of Problem~\ref{p: lmin}
would not say anything for {\em square matrices}, those with $n=d$.
This problem was recently solved in the work \cite{RV}, which 
confirmed prediction \eqref{whp} for general matrices with independent
entries.
Recall that the bounded fourth moment of the entries is necessary 
and sufficient to controll the largest singular value as in \eqref{lmax asymptotic}. 
Then \cite{RV} proves that the fourth moment assumption (i.e. the 
fourth moments of the entries are uniformly bounded) 
is also sufficient to control the smallest singular value.
For an $n \times n$ matrix $A$ with random centered entries of variances
at least $1$, 
\begin{quote}
  {\em Under the fourth moment assumption, prediction \eqref{whp} holds.}
\end{quote}
The identical distribution of the entries is not needed in this result.

For a stronger {\em subgaussian} assumption on the entries, 
prediction \eqref{whp} holds with exponentially high probability.
This was conjectured by Spielman and Teng \cite{ST} for random 
$\pm 1$ matrices:
$$
\P \big( s_n(A) \le \e n^{-1/2} \big) \le \e + c^n,
$$
and proved in \cite{RV} in more generality -- for all matrices
with subgaussian i.i.d. entries, and up to a constant factor
which depends only on the subgaussian moment.

\begin{theorem}[\cite{RV}]
  Let $A$ be an $n \times n$ matrix with i.i.d. centered 
  subgaussian entries with variance $1$.
  Then for every $\e \ge 0$ one has
  \begin{equation}                      \label{eq main}
    \P \big( \lmin(A) \le \e n^{-1/2} \big)
    \le C\e +  c^n,
  \end{equation}
  where $C > 0$  and $c<1$ are constants that depend (polynomially)
  only on the subgaussian moment of the entries.
\end{theorem}

In particular, for $\e = 0$ we deduce that
any random square matrix with i.i.d. subgaussian entries 
with variance $1$ is {\em singular with exponentially small probability}.
For random matrices with $\pm 1$ entries, this was proved by 
Kahn, Koml\'os and Szemer\'edi \cite{KKS}. For more on prior work
and related conjectures on the singularity probability, see \cite{Vu, RV}.


{\small
\begin{thebibliography}{S 99}

\bibitem{Adl83} I. Adler,
  {\em The expected number of pivots needed to solve parametric
  linear programs and the efficiency of the self-dual simplex method}.
  Technical Report, University of California at Berkeley, May 1983

\bibitem{AKS87} I. Adler, R. M. Karp, R. Shamir,
  {\em A simplex variant solving an $m \times d$ linear program
  in $O(\min(m^2,n^2))$ expected number of pivot steps},
  J. Complexity 3 (1987), 372--387

\bibitem{AM85} I. Adler, N. Megiddo,
  {\em A simplex algorithm whose average number of steps is bounded between
  two quadratic functions of the smaller dimension},
  Journal of the ACM 32 (1985), 871--895

\bibitem{Bor} K.-H. Borgwardt,
  {\em The simplex method. A probabilistic analysis.}
  Algorithms and Combinatorics: Study and Research Texts, 1.
  Springer-Verlag, Berlin, 1987.

\bibitem{BN} A. Ben-Tal, A. Nemirovski,
  {\em On polyhedral approximations of the second-order cone},
  Math. Oper. Res. 26 (2001), 193--205

\bibitem{CS} The Compressed Sensing website 
  \verb=http://www.dsp.ece.rice.edu/cs/=

\bibitem {DS} A. Deshpande, D. A. Spielman,
  {\em Improved smoothed analysis of the shadow vertex simplex method},
  FOCS 2005 (46th Annual Symposium on Foundations of Computer Science), 
  349--356

\bibitem{D} D. Donoho,
  {\em Compressed sensing},
  IEEE Trans. Inform. Theory 52 (2006), 1289--1306

\bibitem{E 88} A. Edelman,
  {\em Eigenvalues and condition numbers of random matrices},
  SIAM J. Matrix Anal. Appl.  9  (1988), 543--560

\bibitem{GS} S. Gaas, T. Saaty,
  {\em The computational algorithm for the parametric objective function},
  Naval Research Logistics Quarterly 2 (1955), 39--45

\bibitem {GN}   R.M.Gray, D.L.Neuhoff,
  {\em Quantization},
  IEEE Trans. Inform. Theory  44  (1998), 2325--2383

\bibitem{Hai83} M. Haimovich,
  {\em The simplex algorithm is very good!: On the expected number of
  pivot steps and related properties of random linear programs}.
  Technical report, Columbia University, April 1983

\bibitem{IM} P. Indyk, J. Matou\v sek, 
  {\em Low-distortion embeddings of finite metric spaces},
  in: Handbook of discrete and computational geometry, 177--196.
  Second edition. Edited by Jacob E. Goodman and Joseph O'Rourke. 
  Discrete Mathematics and its Applications (Boca Raton). 
  Chapman \& Hall/CRC, Boca Raton, FL, 2004. 

\bibitem{KKS} J. Kahn, J. Koml\'os, E. Szemer\'edi,
  {\em On the probability that a random $\pm 1$-matrix is singular},
  J. Amer. Math. Soc. 8 (1995), no. 1, 223--240

\bibitem{KK} G. Kalai, D. J. Kleitman,
  {\em A quasi-polynomial bound for the diameter of graphs of polyhedra},
  Bulletin Amer. Math. Soc. 26 (1992), 315--316

\bibitem{KS} J. A. Kelner, D. A. Spielman,
  {\em  A randomized polynomial-time simplex algorithm for Linear Programming},
  submitted.

\bibitem{KMe} B. Klartag, S. Mendelson, 
  {\em Empirical Processes and Random Projections},
  Journal of Functional Analysis 225 (2005), 229--245

\bibitem{KM} V. Klee, G. J. Minty, 
  {\em How good is the simplex algorithm?}
  In Sisha, O., editor, {\em Inequalities -- III}, pp. 159--175, 
  Academic Press, 1972

\bibitem{LV} Yu. Liubarskii, R. Vershynin, 
  {\em Uncertainty principles and vector quantization}, submitted

\bibitem{LPRT} A. Litvak, A. Pajor, M. Rudelson, N. Tomczak-Jaegermann,  
  {\em Smallest singular value of random matrices and geometry of 
  random polytopes},  
  Adv. Math.  195 (2005), 491--523

\bibitem{Meg86} N. Megiddo, 
  {\em Improved asymptotic analysis of the average number of steps 
  performed by the self-dual simplex algorithm},
  Math. Programming  35  (1986), 140--172.

\bibitem{MPT} S. Mendelson, A. Pajor, N. Tomczak-Jaegermann, 
  {\em Reconstruction and subgaussian operators}, 
  Geometric and Functional Analysis, to appear

\bibitem {R} M. Rudelson, 
  {\em Lower estimates for the singular values of random matrices}, 
  Compt. Rendus Math. de L'Academie des Sciences  342  (2006), 247--252.

\bibitem{RV} M. Rudelson, R. Vershynin, 
  {\em The Littlewood-Offord Problem and invertibility of random matrices},
  submitted

\bibitem{SST} S. Sankar, D. Spielman, S.-H. Teng, 
  {\em Smoothed analysis of the condition numbers and growth factors of matrices}, 
  SIAM J. Matrix Anal. Appl. 28 (2006), 446--476

\bibitem{Sch} A. Schrijver,
  {\em Theory of linear and integer programming}.
  Wiley-Interscience Series in Discrete Mathematics. 
  A Wiley-Interscience Publication. John Wiley \& Sons, Ltd., 
  Chichester, 1986

\bibitem{Si} J. W. Silverstein, 
  {\em On the weak limit of the largest eigenvalue of a large-dimensional 
  sample covariance matrix},
  J. Multivariate Anal.  30  (1989), 307--311

\bibitem{Sma1} S. Smale, 
  {\em On the average number of steps of the simplex method of linear programming},
  Math. Programming  27  (1983), 241--262. 

\bibitem{S 85} S. Smale,
  {\em On the efficiency of algorithms of analysis},
  Bull. Amer. Math. Soc. (N.S.)  13  (1985), 87--121

\bibitem{Sma2} S. Smale,
  {\em The problem of the average speed of the simplex method}. 
  Mathematical programming: the state of the art (Bonn, 1982), 
  530--539, Springer, Berlin, 1983.

\bibitem {ST} D. A. Spielman, S.-H. Teng,
  {\em Smoothed analysis: why the simplex algorithm usually takes polynomial time},
  Journal of the ACM 51 (2004), 385--463

\bibitem{ST Congress} D. A. Spielman, S.-H. Teng,
  {\em Smoothed analysis of algorithms},
  Proceedings of the International Congress of Mathematicians,
  Vol. I (Beijing, 2002),  597--606, Higher Ed. Press, Beijing, 2002.

\bibitem{Tod86} M. J. Todd,
  {\em Polynomial expected behavior of a pivoting algorithm
  for linear complementarity and linear programming problems},
  Mathematical Programming 35 (1986), 173--192

\bibitem{Vem1} S. Vempala,
  {\em The random projection method},  
  Handbook of randomized computing, Vol. I, II,  651--671, 
  Comb. Optim., 9, Kluwer Acad. Publ., Dordrecht, 2001.

\bibitem{Vem2} S. Vempala, 
  {\em The random projection method}. 
  With a foreword by Christos H. Papadimitriou. 
  DIMACS Series in Discrete Mathematics and Theoretical Computer Science, 65.
  American Mathematical Society, Providence, RI, 2004.

\bibitem{V} R. Vershynin, 
  {\em Beyond Hirsch Conjecture: walks on random polytopes and smoothed 
  complexity of the simplex method}, FOCS 2006 
  (47th Annual Symposium on Foundations of Computer Science), 133--142

\bibitem{vN} J. von Neumann,
  {\em Collected works.
  Vol. V: Design of computers, theory of automata and numerical analysis}.
  General editor: A. H. Taub. A Pergamon Press Book The Macmillan Co.,
  New York 1963

\bibitem{Vu} V. Vu, 
  {\em Random discrete matrices}, submitted

\bibitem{YBK} Y. Q. Yin, Z. D. Bai, P. R. Krishnaiah, 
  {\em On the limit of the largest eigenvalue of the large-dimensional 
  sample covariance matrix},
  Probab. Theory Related Fields  78  (1988), 509--521

\end{thebibliography}
}
\end{document}